
\input harvmac
\def\eqnz{\eqno(\secsym\the\meqno)\global\advance\meqno by 1}
\def\tr{{\rm tr\,}}

\def\d{{\rm d}}

\def\newsec#1{\global\advance\secno by1\message{(\the\secno. #1)}
\bigbreak\noindent{\it\the\secno. #1}\par\nobreak\smallskip\nobreak}
\input mssymb
%
\noblackbox
\Title{\vbox{\baselineskip12pt\hbox{LPTENS 92/19}\hbox{SPhT/92-064}}}
{Renormalization Group Approach to Matrix Models}

\centerline{Edouard Br\'ezin$^1$  and Jean Zinn-Justin$^2$}
\bigskip{\baselineskip14pt\centerline{${}^1$Laboratoire de Physique
Th\'eorique de l'Ecole Normale Sup\'erieure$^*$}
\centerline{24 rue Lhomond, 75231 Paris Cedex 05, France}
\centerline{E-mail:brezin@physique.ens.fr}}
\medskip
{\baselineskip14pt\centerline{${}^2$Service de Physique
Th\'eorique$^\dagger$}
\centerline{CE-Saclay, 91191 Gif-sur-Yvette Cedex, France}
\centerline{E-mail:zinn@poseidon.saclay.cea.fr}}
\footnote {}{$^*$Unit\'e
propre du Centre National de la Recherche Scientifique, associ\'ee \`a
l'Ecole Normale Sup\'erieure et \`a l'Universit\'e Paris-Sud}
\footnote{}{$^{\dagger}
$Laboratoire de la Direction des Sciences de la Mati\`ere du Commissariat
\`a l'Energie Atomique}
\bigskip
{\it Abstract.} Matrix models of 2D quantum gravity are either exactly
solvable for matter of central charge $ c\leq 1, $ or not understood. It
would be useful to devise an approximate scheme which would be
reasonable for the known cases and could be carried to the
unsolved cases in order to achieve at least a qualitative
understanding of the properties of the models. The double scaling
limit is an indication that a change of the length scale induces
a flow in the parameters of the theory, the size of the matrix
and the coupling constants for matrix models, at constant long
distances physics. We construct here these renormalization group
equations at lowest orders in various cases to check that we
reproduce qualitatively the properties of $ c\leq 1 $ models.
\Date{06/92}

\newsec{Introduction}
The matrix model representation of two-dimensional
quantum gravity \ref\rDKAJ{F. David, {\sl Nucl. Phys.\/} {\bf B311}
(1985) 45 and 543\semi
V. Kazakov, {\sl Phys.Lett.\/} {\bf 150B} (1985) 28\semi
J. Ambj\o rn, B. Durhuus and J. Fr\"ohlich, {\sl Nucl. Phys.\/} {\bf B257}
(1985) 433.} has led to explicit solutions for
minimal
conformal fields couped to gravity \ref\rBKDSGM{E. Br\'ezin and V. Kazakov,
{\sl Phys. Lett.\/} {\bf B236 }(1990) 144\semi
M. Douglas and S. Shenker, {\sl Nucl. Phys.\/} {\bf B235} (1990) 635\semi
D.G. Gross and A.A. Migdal, {\sl Phys. Rev. Lett.\/} {\bf 64} (1990) 127.}
and to beautiful connections with various branches of mathematical physics
such as integrable flows or topological field theories. However this
approach has not been of real help for understanding the
difficulties which arise when the central charge $ c $ of the matter
field is larger than one although it is very easy to write matrix
models for those cases as well. Indeed if the topological
expansion of field theories in dimension larger than one is
notoriously difficult to handle, one can write (for instance) a
matrix model for $ n $ Ising spins coupled to a randomly triangulated
surface, with a central charge $ c=n/2, $ as an integral over $ 2^n $
matrices \ref\rBHik{E. Br\'ezin and S. Hikami, ``A Naive
Matrix-Model Approach to
Two-Dimensional Quantum Gravity Coupled to Matter of Arbitrary
Central Charge", Ecole Normale Sup\'erieure, Preprint LPTENS 92/10.}.
Unfortunately these models are not solvable (up to
now) for $ n>1. $ \par
Therefore we are in the unfortunate situation of either
solving exactly (for $ c \leq  1) $ or not understanding at all what is
happening. We are not even guaranteed that a matrix model
candidate to describe a $ c>1 $ model has anything to do with the
continuum description, although we should keep in mind that (i) it
is not yet clear that there is a continuum theory at present for
$ c>1 $, (ii) we have invested enough efforts in matrix models to try
to find out whether they do have a continuum limit for $ c>1 $; is
there still a double scaling limit with some new type of
exponent, or no continuum limit at all? If there is a continuum
limit is it still a string theory in non-critical dimension? \par
We have no answer to provide to these important
questions, but we would like to devise a method which may be only
approximate for the soluble cases $ c \leq  1 $, but which would allow us
to understand at least qualitatively $ c>1 $. The attempt that we
present here is based on simple minded analogies with critical
phenomena. We note first that the central result of matrix models
is the existence of the ``double scaling limit" \rBKDSGM, i.e.\ a continuum
limit with critical exponents which describe how the two coupling
constants of the theory have to be tuned to reach this limit.
These simple scaling laws for $ c<1 $, with logarithmic deviations at
$ c=1$ \ref\rBKZGP{
E. Br\'ezin, V. Kazakov and Al. A. Zamolodchikov, {\sl Nucl. Phys.\/}
{\bf B338} (1990) 673\semi
P. Ginsparg and J. Zinn-Justin, {\sl Phys. Lett.\/} {\bf B240} (1990)
333\semi
D.J. Gross and N. Miljkovic, {\sl Phys. Lett.\/} {\bf B238} (1990) 217\semi
G. Parisi, {\sl Phys. Lett.\/} {\bf B238} (1990) 209.} are reminiscent of the
theory of phase transitions for  dimensions larger or equal to four. There we
know that a  non-trivial IR fixed point governs dimensions smaller
than  four and that we have to rely on approximate methods such as
the$\varepsilon$  or $ 1/N $ expansions. Is there a similar phenomenon for
matrix  models? This question makes sense only if we understand
renormalization group flows. Indeed if we had such flow
equations, we know that the scaling laws and the exponents, which
characterize the double scaling limit, would arise automatically.
Therefore we have to understand how the two coupling constants of
the theory, the string coupling constant and the cosmological
constant (mapped respectively into the size $ N $ of the matrices and
the matrix coupling constant $ g $), evolve under a rescaling of the
regularization length introduced in the triangulation of the
worldsheet. We thus expect that a change $ N \mapsto N+\delta N $ can
be compensated by a change $ g \mapsto g+\delta g $ with the same
continuum physics. We shall see that for matrix models this indeed the case,
at the expense of enlarging the space of coupling constants very much as
in the Wilson's scheme \ref\rWK{K.G. Wilson and J. Kogut, {\sl Phys.
Rep.\/} {\bf 12C} (1974) 75.} of integration over the momenta in the
shell $ \Lambda - \d  \Lambda  < \vert p\vert  < \Lambda  $. \par
Let us review first briefly the $ c<1 $ scaling laws \ref\rKPZ{V.G. Knizhnik,
A.M. Polyakov and A.A. Zamolodchikov, {\sl Mod.
Phys. Lett.\/} {\bf A3} (1988) 819\semi
F. David, {\sl Mod. Phys. Lett.\/} {\bf A3} (1988) 207\semi
J. Distler and H. Kawai, {\sl Nucl. Phys.\/} {\bf B231} (1989) 509.}:
the  singular part of the string partition function satisfies
\eqn\escalun{Z = \Delta^{ 2-\gamma_ 0}f \left(\Delta \ N^{2/\gamma_ 1}
\right) }
with
$$ \Delta = g_c - g \eqnz $$
and
$$ \gamma_ 1 = 2 - \gamma_ 0 = {1\over12} \left[25-c+ \sqrt{( 1-c)(25-c)}
\right] . \eqnz $$
The string susceptibility exponent at genus $ h $ is
$$ \gamma_ h = \gamma_ 0 + h\, \gamma_ 1\,. \eqnz $$
The relation
\eqn\egamrel{\gamma_ 0+\gamma_ 1=2 }
independent of the explicit values of $ \gamma_ 0 $ and $ \gamma_ 1 $, is
easily obtained from the consideration of the torus. \par
Assume now that we have indeed a Callan--Symanzik like
differential equation for this string partition function which
reads
\eqn\eZRG{[N\partial /\partial N - \beta( g)\partial /\partial g + \gamma(
g)] Z(N,g) = r(g)}
with a fixed point $ g^\ast $ given by $ \beta \left(g^\ast \right) = 0 $ and
$ \beta^{ \prime} \left(g^\ast \right)>0 $. It is then
elementary to verify that we recover the scaling law \escalun\ with
$$ \gamma_ 1 = {2 \over \beta^{ \prime} \left(g^\ast \right)}\ ,\qquad
\gamma_ 0 = 2 - {\gamma \left(g^\ast \right) \over \beta^{ \prime}
\left(g^\ast \right)}. \eqnz $$
With several coupling constants $ g_k $, the scaling exponents are
given by the eigenvalues of the matrix
\eqn\eOM{\Omega_{ kl} = {\partial \beta_ k \over \partial g_l} \left(g^\ast_ m
\right) .}
Note that the scaling law \egamrel\ requires that $ \gamma \left(g^\ast
\right)=2 $.
\par
These flow equations will be constructed by integrating out
one line and one row of an $ (N+1)\times( N+1) $ matrix, thereby reducing
it to an $ N\times N $ matrix. In this process we shall prove that the
matrix partition function $ \zeta_ N(g) $ fulfills the equation
$$ \zeta_{ N+1}(g) = [\lambda( g)]^{N^2}\zeta_ N \left(g^{\prime} \right)
\eqnz $$
with
$$ g^{\prime}  = g + {1 \over N} \beta( g) + O \left({1 \over N^2} \right)
\eqnz $$
and
$$ \lambda( g) = 1 + {1 \over N} r(g) + O \left({1 \over N^2} \right). \eqnz
$$
It follows immediately that the string partition function
$$ Z(N,g) = {1 \over N^2} \ln\zeta_ N(g) \eqnz $$
satisfies \eZRG\ with
$$ \gamma( g) = 2 \,,\eqnz $$
and therefore the scaling law \egamrel\ does hold. \par
The set-up of this article is the following: we first
discuss at lowest order pure gravity, then multicritical points
of one-matrix models, then a more appropriate saddle-point method
and the new features which manifest themselves at higher orders.
\newsec{A Simple Perturbative Calculation}
We begin with a one-matrix $ \phi^ 4 $ model, which near its
critical point describes pure gravity $ (c=0) $. It consists of an
integral over an $ N\times N $ hermitian matrix $ \phi_ N $:
$$ \zeta_ N(g) = \int^{ }_{ } \d  \phi_ N\, \exp\left[- S_N \left(\phi_N,g
\right)\right], \eqnz $$
with an action
\eqn\eactSN{S_N \left[\phi_ N,g \right] = N\, \tr \left[{1 \over 2}\phi^
2_N+{g \over 4}\phi^ 4_N \right].}
{}From the exact solution \rBKDSGM\ we know that the double scaling
limit
is reached in the vicinity of
$$ g_c = -1/12 \eqnz $$
with an exponent
$$ \gamma_ 1 = 5/2\,. \eqnz $$
The matrix $ \phi_{ N+1} $ is parametrized in terms of an $ N\times N $
submatrix $ \phi_ N, $
a complex $ N $-component vector $ v_a, $ and a number $ \alpha$:
\eqn\ephiNun{\phi_{ N+1} = \left( \matrix{\displaystyle \phi_ N &
\displaystyle v_a
\cr\displaystyle v^\ast_a & \displaystyle \alpha \cr} \right) ,}
but one verifies easily that all the terms involving $ \alpha $ are of
relative order $ 1/N $ and can be dropped in the continuum limit; in
other words we can set $ \alpha =0 $. We then define
$$ \exp\left[- S^{\prime}_ N \left(\phi^{ \prime}_ N,g^{\prime}\right)\right]
= \lambda_ N(g) \int \d^Nv\, \d^Nv^\ast \, \exp\left[- S_{N+1}
\left(\phi_ N,v,g \right)\right], \eqnz $$
in which $ \phi^{ \prime}_ N $ is obtained after a rescaling which normalizes
the
coefficient of $ \phi^ 2_N $ to $ N/2 $ as in \eactSN. \par
An easy calculation yields
\eqn\eSNun{S_{N+1} \left(\phi_{ N+1} \right) = (N+1) \left[ \tr \left({1 \over
2}\phi^ 2_N + {g \over 4} \phi^ 4_N \right)+v^\ast \cdot v \right]+(N+1)g
\left[v^\ast \phi^ 2_Nv+{1 \over 2} \left(v^\ast ,v \right)^2 \right] ,}
after which we expand the exponential to first order in $ v, $ perform the
easy integrations over the $ v_a$'s and re-exponentiate to get
a new effective action
$$ S^{\prime} \left(\phi_ N \right) = (N+1) \left[ \tr \left({1 \over 2}
\phi^ 2_N+{g \over 4}\phi^ 4_N \right) \right]+g \tr \phi^ 2_N\,. \eqnz $$
We then rescale $ \phi_ N $ to
$$ \phi_ N = \rho  \phi^{ \prime}_ N \eqnz $$
with
$$ \rho  = 1 - {2g+1 \over 2N} + O \left({1 \over N^2} \right) \eqnz $$
(we are dropping from now on all terms of relative order $ 1/N^2 $ as well as
higher orders in $ g $). This rescaling gives as coefficient of $ \phi^ 4 $
$$ g^{\prime}  = g - {1 \over N} \left(g+4 g^2 \right) \eqnz $$
and for the prefactor in the partition function
$$ \lambda( g) = 1 - {1 \over N}(1+3 g). \eqnz $$
Therefore we have established at first order the renormalization group
equation \eZRG\ with
\eqna\eRGfun
$$ \eqalignno{ \beta( g) & = - g - 4 g^2 + O \left(g^3 \right) & \eRGfun{a}
\cr \gamma( g) & = 2 & \eRGfun{b} \cr r(g) & = - 1 - 3 g + O \left(g^2
\right). & \eRGfun{c} \cr} $$
There are two fixed $ g^\ast  = 0 $ or $ -1/4 $, but we have to select the
repulsive fixed point
$$ g^\ast  = -1/4 \eqnz $$
since the pure gravity exponents are obtained only when we ``tune" the
cosmological constant $ g $ near its critical value $ g_c $; the true $ g_c $
is $ -1/12 $ in the exact theory instead of our first approximation $ -1/4 $.
Since at this order $ \beta^{ \prime} \left(g_c \right) $ is equal to one,
this calculation gives
$$ \gamma_ 1 = 2\,, \eqnz $$
to be compared to the exact value $ 5/2 $. \par
We shall return to this $ c=0 $ case in a more elaborate calculation
below, but we first consider within the same approximation the
multicritical points of one-matrix models.
\newsec{Multicritical Points}
In order to study the existence of Kazakov's multicritical
points \ref\rKaza{V. Kazakov, {\sl Mod. Phys. Lett.\/}
{\bf A4} (1989) 1691.}  within one-matrix models, we allow for an
arbitrary polynomial in the action
$$ S \left[N,g_k \right] = N \sum^{ \infty}_ 1{g_k \over 2k} \tr
\phi^{ 2k} \eqnz $$
with $ g_1=1 $. \par
The parametrization \ephiNun\ of $ \phi_{ N+1} $ gives a number of terms,
since
$$ \tr  \phi^{ 2k}_{N+1} = \tr \phi^{ 2k}_N + 2k v^\ast \phi^{2k-2}_Nv+O
\left(v^4 \right) .\eqnz $$
The crudest calculation consists in keeping these quadratic terms in $ v $,
expanding to first order in the $ g_k$'s, integrating over the $ v$'s and
reexponentiating. This gives, up to a normalization constant,
$$ \zeta_{ N+1} = \int\d  \phi_ N\, \exp\left\{  -(N+1) \tr
\left[ \left({1 \over 2} + {g_2 \over N+1} \right)\phi^ 2_N + \sum^{ }_ 2
\left({g_k \over 2k} + {g_{k+1} \over N+1} \right)\phi^{ 2k}_N
\right]\right\}  .\eqnz  $$
Rescaling $ \phi $ in order to set the coefficient of $ \phi^ 2_N $ to $ N/2,
$ we obtain
$$ g^{\prime}_ k = g_k + {1 \over N} \left[- \left(2k g_2 + k-1
\right)g_k+2k g_{k+1} \right] ,\eqnz $$
and thus
$$ \beta_ k = - \left(2k g_2+k-1 \right) g_k +2k g_{k+1}. \eqnz $$
It is easy to verify that there is a multicritical point of order $ m $
given by
$$\left. g^\ast_ k = \cases{\displaystyle\left(-1 \over 2m\right)^{k-1}
{m-1\choose k-1}  & \qquad for $ 2\leq k\leq m $ \cr
0 & \qquad for \quad $m <k $. \cr} \right. \eqnz $$
The corresponding multicritical potential is
$$ S^\ast_ N = N\, \tr \left[1- \left(1 - {\phi^ 2_N \over 2m} \right)^m
\right] \eqnz $$
which is of course approximate; for $ m=2 $ it gives $ x^2/2 - x^4/16 $,
instead
of the exact \rBKDSGM\ $ x^2/2 - x^4/48 $; for $ m=3 $, $
x^2/2-x^4/12 + x^6/216 $ instead of $ x^2/2 - x^4/12 + x^6/180 $.
Although it is only approximate, we do find the right sequence of
multicritical points with their characteristic features of being
alternatively unbounded below for even $ m$'s and bounded for
odd $ m$'s. \par
The critical exponents are related to the matrix \eOM\ of
derivatives of the $ \beta $-functions at the fixed points. For the $ m $-th
multicritical points this matrix is real, $ (m-1)\times( m-1) $ and
non-symmetric.
Remarkably enough this matrix turns out to have $ (m-1) $ real eigenvalues
which are $ 2/m $, $ 3/m$,...,$m/m$. For an arbitrary direction of approach
of the fixed point the leading critical exponent is the largest one, namely
one, which corresponds in fact to the $ c=0 $ exponent. If we choose a
direction in the space spanned by the $ (m-2) $ subleading eigenvalues we
shall see in general as leading exponent $ (m-1)/m $. In order to tune the
$ m $-th multicritical point we have to suppress the $ (m-2) $ largest
eigenvalues by choosing a direction in the $ g_k-g^\ast_ k $ space along the
eigenvector of the smallest eigenvalue, namely $ 2/m $. This gives for the
$ m $-th multicritical point
$$ \gamma_ 1=m \eqnz $$
whereas the exact value is $ m+1/2 $. This trivial calculation is thus in
good qualitative and semi-quantitative agreement with the exact answer.
\newsec{A Better Calculation}
An attempt to expand to higher orders in the coupling constant
in order to improve the above calculations reveals that we have not done
it in a systematic way. For instance if we follow the same procedure to
next order and try a ``two-loop" calculation, we find that the
$ \beta $-function at order $ g^2 $ receives contributions from higher orders.
In fact we have to integrate first over the $ N $ complex $ v $'s without
perturbation theory, and then expand in order to reach the large $ N $,
double-scaling limit. To this effect we write the effective action
$$\left.\eqalign{ S_{N+1} \left[\phi_ N,v_a,\sigma \right] &
= (N+1) \left[ \tr \left({1 \over 2}\phi^ 2_N + {g \over 4}
\phi^ 4_N \right)+v^\ast \cdot v \right]  \cr  &
\quad + (N+1) g \left[v^\ast \phi^ 2_Nv + \sigma  v^\ast \cdot v -
{\sigma^ 2 \over 2} \right] , \cr}\right. \eqnz $$
which, upon integration over $ \sigma$, is equivalent to \eSNun. We next
integrate over the $ v $'s and obtain
$$ S^{ {\rm eff}}_N \left[\phi_ N,\sigma \right] = (N+1) \left[ \tr
\left({1 \over 2} \phi^ 2_N + {g \over 4} \phi^ 4_N \right) \right] - (N+1)
{g\sigma^ 2 \over 2} + \tr  \ln \left(1+g \sigma +g \phi^ 2_N
\right). \eqnz $$
In the large $ N $ limit, the single $ \sigma $-mode is fixed to its
saddle-point value given by
$$ \sigma  = {1 \over N} \tr \left(1+g \sigma +g \phi^ 2_N \right)^{-1}.
\eqnz $$
This gives $ \sigma $ as a functional of the traces of the even powers of $
\phi $, involving also non-linear terms as $ \left( \tr  \phi^ 2_N \right)^2
$ and so on.
Therefore the original action has to be enlarged to accomodate such terms.
However is we expand again in powers  of $ g $ these new interactions do not
yet appear at order $ g^2 $ since:
$$ \sigma  = 1 - \left(1+t_2 \right) g + \left(2+3 t_2+t_4 \right)+O
\left(g^3 \right) \eqnz $$
in which
$$ t_k = {1 \over N} \tr  \phi^ k \eqnz $$
and thus, dropping all terms of order $ g^3 $
$$ {S^{ {\rm eff}} \over N} = \left[{1 \over 2} (N+1) + g - g^2 \right] t_2 +
\left[{g \over 4} (N+1) - {g^2 \over 2} \right] t_4\,. \eqnz $$
After a rescaling of $ \phi_ N $ to enforce to $ N/2 $ the coefficient of $
t_2 $ we find a new $ \phi^ 4 $ coupling constant
$$ g^{\prime}  = g - {1 \over N} \left(g+6 g^2 \right)+O \left(g^3 \right)
\eqnz $$
i.e.
$$ \beta( g) = -g - 6 g^2 + O \left(g^3 \right). \eqnz $$
The fixed point $ g^\ast  = -1/6 $ is a slight improvement over the previous
calculation, but at this order the exponent $ \gamma_ 1 $ does not change.
\par
A calculation at next order requires two additional coupling
constants, namely that of $ \tr  \phi^ 6 $ and of $ \left( \tr \
\phi^ 2 \right)^2 $; the renormalization flow
takes place in a three dimensional space. The result of this calculation will
be reported elsewhere.
\newsec{Conclusion}
Simple renormalization group transformations reproduce qualitatively
and semi-quantitatively the results of matrix models. In principle we could
apply readily the same technique to matrix models with $ c $ larger than one.
However manifestly the method has first to pass a few non-trivial tests: can
it be systematically improved at higher orders for simple one-matrix
models?
For $ c=1 $ we should find that the fixed point is a double zero of the $
\beta $-function in order to account for the logarithmic deviations to
scaling \rBKZGP.
The situation is still unclear, but we believe that it is worth exploring the
possibilities of this approach further.
\listrefs
\bye
\vfill\eject
\centerline{{\bf REFERENCES}}
\bigskip
\noindent \item {$\lbrack$1$\rbrack$}F. David, {\sl Nucl. Phys.\/} {\bf B311}
(1985) 45 and 543; \par
\noindent \item {\nobreak\ \nobreak\ \nobreak\ }V. Kazakov, {\sl Phys.
Lett.\/} {\bf 150B} (1985) 28; \par
\noindent \item {\nobreak\ \nobreak\ \nobreak\ }J. Ambj\o rn, B. Durhuus and
J. Fr\"ohlich, {\sl Nucl. Phys.\/} {\bf B257}
(1985) 433. \par
\noindent \item {$\lbrack$2$\rbrack$}E. Br\'ezin and V. Kazakov, {\sl Phys.
Lett.\/} {\bf B236 }(1990) 144; \par
\noindent \item {\nobreak\ \nobreak\ \nobreak\ }M. Douglas and S. Shenker,
{\sl Nucl. Phys.\/} {\bf B235} (1990) 635; \par
\noindent \item {\nobreak\ \nobreak\ \nobreak\ }D.G. Gross and A.A. Migdal,
{\sl Phys. Rev. Lett.\/} {\bf 64} (1990) 127. \par
\smallskip
\noindent \item {$\lbrack$3$\rbrack$}E. Br\'ezin and S. Hikami, \lq\lq A Naive
Matrix-Model Approach to
Two-Dimensional Quantum Gravity Coupled to Matter of Arbitrary
Central Charge\rq\rq , Ecole Normale Sup\'erieure, Preprint LPTENS 92/10. \par
\smallskip
\noindent \item {$\lbrack$4$\rbrack$}K.G. Wilson and J. Kogut, {\sl Phys.
Rep.\/} {\bf 12C} (1974) 75. \par
\smallskip
\noindent \item {$\lbrack$5$\rbrack$}V.G. Knizhnik, A.M. Polyakov and A.A.
Zamolodchikov, {\sl Mod.
Phys. Lett.\/} {\bf A3} (1988) 819; \par
\noindent \item {\nobreak\ \nobreak\ \nobreak\ }F. David, {\sl Mod. Phys.
Lett.\/} {\bf A3} (1988) 207; \par
\noindent \item {\nobreak\ \nobreak\ \nobreak\ }J. Distler and H. Kawai, {\sl
Nucl. Phys.\/} {\bf B231} (1989) 509. \par
\smallskip
\noindent \item {$\lbrack$6$\rbrack$}V. Kazakov, {\sl Mod. Phys. Lett.\/}
{\bf A4} (1989) 1691. \par
\smallskip
\noindent \item {$\lbrack$7$\rbrack$}E. Br\'ezin, V. Kazakov and Al. A.
Zamolodchikov, {\sl Nucl. Phys.\/}
{\bf B338} (1990) 673; \par
\noindent \item {\nobreak\ \nobreak\ \nobreak\ }P. Ginsparg and J.
Zinn-Justin, {\sl Phys. Lett.\/} {\bf B240} (1990)
333; \par
\noindent \item {\nobreak\ \nobreak\ \nobreak\ }D.J. Gross and N. Miljkovic,
{\sl Phys. Lett.\/} {\bf B238} (1990) 217; \par
\noindent \item {\nobreak\ \nobreak\ \nobreak\ }G. Parisi, {\sl Phys. Lett.\/}
{\bf B238} (1990) 209. \par
\smallskip
\bye